# ARTICLE

# Nano-particle motion in monolithic silica column using single-particle tracking method

Yusaku Abe[*,a], Naoki Tomioka[a] and Yu Matsuda[*,a]

Porous materials are used in a variety of industrial applications owing to their large surface areas, large pore volumes, hierarchical porosities, and low densities. The motion of particles inside the pores of porous materials has attracted considerable attention. We investigated nano-particle motion in a porous material using the single-particle tracking method. Particle motion such as absorption and desorption at the wall was observed. The displacement probability distribution deviated from the Gaussian distribution at the tail, indicating non-Gaussian motion of the particles. Moreover, an analysis of the relative angle between three consecutive particle positions revealed that the probability of the particle moving backward was approximately twice that of the particle moving forward. These results indicate that particle motion inside porous materials is highly complex and that a single-particle study is essensital for fabricating a structure that is suitable for applications.

## Introduction

Porous materials have been used as filter, absorbent, catalyst, heat exchanger, and pump owing to their unique properties such as their large surface area, large pore volume, hierarchical porosity, and low density. [1-4] The characteristics of porous materials have been investigated by imaging techniques such as X-ray computed tomography (CT), energy dispersive X-ray spectroscopy (EDS), scanning electron microscopy (SEM), transmission electron microscopy (TEM), mercury porosimetry, and optical microscope. [5-7] These techniques reveal the structures (pore size and tortuosity factor) and elemental maps of porous materials. As particle motion in porous materials is important for the design of efficient devices, studies on particle motion in porous materials have been conducted. Several theoretical and simulation-based studies have been conducted on particle motion in idealised geometry channels. [8-12] Experimental approaches are also effective for revealing particle motion in porous materials. The diffusion behaviour of particles in porous materials has been investigated using dynamic light scattering (DLS) [13-15] and fluorescent correlation spectroscopy (FCS). [16, 17] These methods reveal the confined/suppressed diffusion of particles inside pores. However, the measurements are limited to ensemble averaged behaviour. Recently, single-molecule fluorescence imaging method was used to investigate particle diffusion in porous materials. [18, 19] In particular, single-particle/molecule tracking (SPT/SMT), [20-22] in which individual molecule/particle diffusion motions are detected using an optical microscope, has received considerable attention. Particle motions in well-defined straight/curved channels were visualised using the SMT method. [23-25] The particle motions in metal-organic frameworks (MOFs) were investigated, and heterogeneous diffusion motions were observed. [26] The translational and orientational diffusion coefficients of dye molecules with different lengths and charges in surfactant-templated mesoporous silica pores (cylindrical nanoscale pores) were also measured by SMT. [27] Whereas the diffusion motions were similar regardless of the charge of the dye molecules, the diffusion coefficients depended on the interaction between the charge of the dye molecules and the surfactant. For a more complex geometry, the resistance time, diffusion coefficient, and spatial distribution of particles in porous silica material were investigated. [28] The SMT method can observe individual particle motions; for example, the resistance time of a particle with a stuck event was found to be 10 times longer than that of a particle without a stuck event. On the other hand, the particle motions inside pores have not been investigated. The diffusion motions of fluorescent molecules in a fluid catalytic cracking (FCC) particle were observed, and the spatial diffusion coefficient distribution was obtained by SMT. [29] The dye molecules exhibited three diffusion modes: immobile (88% of the dye molecules), mobile (8%), and hybrid motion of immobile and mobile (4%). However, the analysis was limited to the diffusion coefficient.

In the present study, we investigated the nano-particle behaviour in a monolithic silica column with a bimodal pore size distribution consisting of mesopores and macropores [30] using the SPT method. The column is used for practical applications such as liquid chromatography, antibody purification, and DNA purification. The structure of the pores was highly complex, but the pore size was well-controlled. Individual particle motions in the pores were directly observed and analysed. The diffusion mode and direction of particle motion for each particle were

a. Department of Modern Mechanical Engineering, Waseda University, 3-4-1 Ookubo, Shinjuku-ku, Tokyo, 169-8555, Japan
*Corresponding author: Y.A.: ya-jupiter0309@toki.waseda.jp and Y.M.: y.matsuda@waseda.jp

investigated. Moreover, the displacement distribution was calculated to determine whether the diffusion was Gaussian.

## Materials and methods

### Sample preparation

A monolithic silica column (Ex-Pure, Kyoto Monotech Co. Ltd., Japan) was used as received without surface modification. Prior to the measurements, the column characteristics were confirmed. Fig. 1a shows a SEM image of the column. As shown in Fig. 1a, many continuous pores, which are formed by polymerisation-induced phase separation,[30] are observed. The pore size distribution of the column was measured by a mercury porosimeter (AutoPore IV 9500, Shimadzu CO. Ltd., Japan) as shown in Fig. 1b. The column had a bimodal pore size distribution of 52 nm (mesopore) and 2 μm (macropore). The porosity and tortuosity of the column were measured as 83.2% and 4.04, respectively. We employed cadmium-selenium quantum dots (QDs) (Qdot 545, Invitrogen, Thermo Fisher Scientific, USA) as probe particles in the SPT measurement. The manufacturer reported that the diameter of the QD was 20 nm. We prepared a QD toluene solution with a concentration of $1 \times 10^{-10}$ M. The monolithic silica column was immersed in the QD solution in a custom-made glass cell with a diameter and depth of 6 mm and 5 mm, respectively. Both the bottom and cover glass plates (Cover glass, Thickness No.1, Matsunami Glass Co. Ltd., Japan) of the cell were cleaned using an oxygen plasma cleaner (PC-400T, STREX, Inc., Japan).

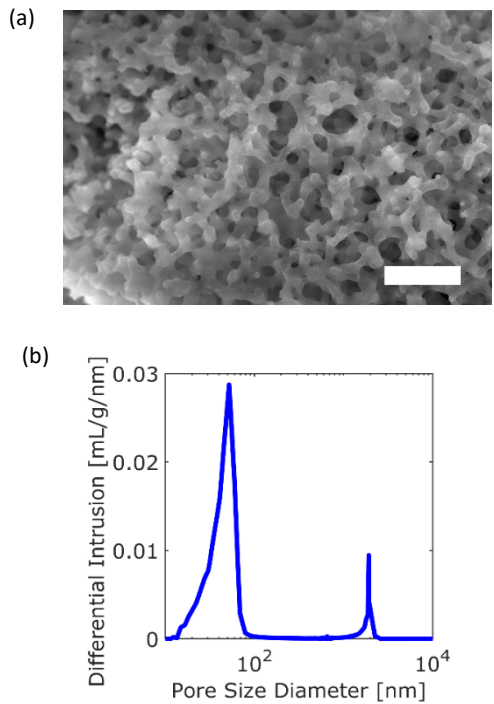

Fig. 1 Characteristics of monolithic silica column
(a) SEM image of the column (scale bar = 10 μm).
(b) Pore size distribution.

### SPT instrumentation

SPT studies were conducted using the same apparatus employed in our previous study,[31] with some modifications. A brief explanation including the modifications is provided below. The SPT experiments were conducted at room temperature. We used an inverted fluorescence microscope (IX-73, Olympus Co. Ltd., Japan) with an oil-immersion objective lens of 100×, NA = 1.45, and WD = 0.13 mm (UPLXAPO100XO, Olympus Co. Ltd., Japan) and a confocal scanner unit (CSU-X1, Yokogawa Electric Co. Ltd., Japan) with a zoom lens of 2.0×. The QDs were excited by a solid-state laser with an emission wavelength of 488 nm (OBIS488LS, Coherent CO. Ltd., USA). The laser output power was set to 170 mW. The emissions from the QDs were recorded using an electron-multiplying charge-coupled device (EMCCD) camera (C9100-23B, ImagEM X2, Hamamatsu Photonics Co. Ltd., Japan) at 20 frames per second (exposure time: 0.05 s). In this setup, one pixel of the image corresponded to 0.08 μm; thus, the observed image area corresponded to 40.96 μm × 40.96 μm. The focal plane was set 5 μm above the bottom surface of the column using a piezo actuator (P-725K, Physik Instrumente GmbH & Co. KG, Germany).

### Analysis of trajectories

The obtained images were analysed using ParticleTracker,[32-34] which is a plugin of ImageJ Fiji.[35] We extracted trajectories from the images using the following the parameters in ParticleTracker: Radius = 3 px, Cutoff = 0.01, Displacement = 7 px, Link range = 1, and Per/Abs = 0.40. In this study, only trajectories longer than six frames were analysed because shorter trajectories are less reliable. The diffusion coefficient of each particle was calculated by ParticleTracker. The time-averaged mean squared displacement (MSD) $\delta^2(\tau)$ with respect to the lag time $\tau$ is defined as

$$\delta^2(\tau) = \frac{1}{T-\tau} \sum_{t=0}^{T-\tau-1} \|\mathbf{r}(t+\tau) - \mathbf{r}(t)\|^2, \qquad (1)$$

where $\mathbf{r}$ is the position vector of the particle, and $T$ represents the total time of the tracked trajectory. $\|\mathbf{a}\|$ is the Euclidean norm of the vector $\mathbf{a}$. The MSD is also represented using the regular two-dimensional diffusion coefficient $D$ as follows:

$$\delta^2(\tau) = 4D\tau^2. \qquad (2)$$

By calculating the intercept of $\log \delta^2(\tau)$ in the plot of $\log \delta^2(\tau)$ versus $\log \tau$, we can extract the diffusion coefficient $D$ as follows:

$$D = \frac{1}{4} \exp(y_0), \qquad (3)$$

where $y_0$ denotes the intercept. Generally, the $\nu$ moment of displacement $\delta^\nu(\tau)$ can be expressed as

$$\delta^\nu(\tau) = \frac{1}{T-\tau} \sum_{t=0}^{T-\tau-1} \|\mathbf{r}(t+\tau) - \mathbf{r}(t)\|^\nu. \qquad (4)$$

Usually, the following scaling law holds:

$$\delta^\nu(\tau) \propto \tau^\gamma. \qquad (5)$$

The scaling coefficient $\gamma$ can be determined via the least squares method from the obtained trajectory data. The plot of $\gamma$ and $\nu$ is called the moment scaling spectra (MSS), and the slope of the MSS is known as the MSS slope $S_{\text{MSS}}$.[32, 36] The MSS slope $S_{\text{MSS}}$ indicates the diffusion mode of the particle: $S_{\text{MSS}} = $

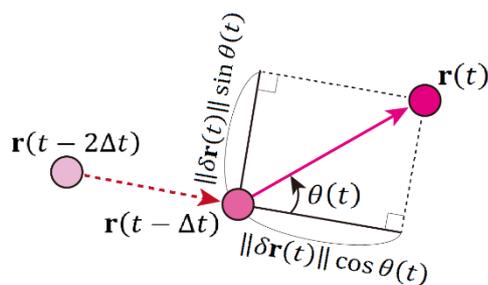

Fig. 2 Schematic of three consecutive particle positions and relative angle.

0.5 for free (normal) diffusion, $0 < S_{\mathrm{MSS}} < 0.5$ for confined or suppressed diffusion, and $0.5 < S_{\mathrm{MSS}} < 1$ for super diffusion including drifted diffusion and Lévy flights. A more detailed explanation is provided in the literature. [32, 36]

The displacement probability distributions (DPDs) of the QDs were calculated from the trajectory data, where the displacements between consecutive frames were calculated for all trajectories as follows:

$$\Delta r = \|\mathbf{r}(t+\Delta t) - \mathbf{r}(t)\| . \quad (6)$$

Here, $\Delta t$ is the frame interval. The DPD for a freely diffusing particle undergoing the Brownian motion is expressed as [37-39]

$$G_s(\Delta r) = \frac{1}{\sqrt{4\pi D \Delta t}} \exp\left(-\frac{\Delta r^2}{4 D \Delta t}\right). \quad (7)$$

One of the advantages of SPT over ensemble methods, such as DLS, is that particle motion can be directly observed as time-series images. Then, the relative angle $\theta$ is calculated from three consecutive position vectors as shown in Fig. 2.[40-42] The relative displacements are defined as $\delta \mathbf{r}(t) = \mathbf{r}(t) - \mathbf{r}(t - \Delta t)$ and $\delta \mathbf{r}(t - \Delta t) = \mathbf{r}(t - \Delta t) - \mathbf{r}(t - 2\Delta t)$. The relative angle $\theta(t)$ is expressed as

$$\theta(t) = \cos^{-1}\left[\frac{\delta \mathbf{r}(t) \cdot \delta \mathbf{r}(t - \Delta t)}{\|\delta \mathbf{r}(t)\| \|\delta \mathbf{r}(t - \Delta t)\|}\right]. \quad (8)$$

When $\theta = 0°$, the particle moves in the same direction as the previous time. On the other hand, when $\theta = 180°$, the particle moves backward. For free diffusion, the probability distribution of the relative angle $\theta$ is uniform.

## Results and discussion

### Spatial distribution of trajectories

Typical SPT data captured over 500 frames (25 s) are shown in Fig. 3, where the trajectories coloured according to their diffusion coefficients are overlayed on the white-light transmission image. The white areas correspond to the pores in the column. All trajectories were recognised in the pore areas. As shown in Fig. 3, the diffusion coefficients were widely distributed over three orders of magnitude. There were immobile QDs (absorbed QDs) near the wall (black area), as shown in Fig. 3c. Freely diffusing QDs were observed at the centre of the pores, as shown in Fig. 3d. A QD trajectory with absorption and desorption was recognised near the bottleneck, as shown in Fig. 3b.

### Statistical behaviour of trajectories

A histogram of the diffusion coefficient calculated by Eq. 3 is shown in Fig. 4a, where the number of bins is determined by the Sturges' rule. This figure also indicates a broad distribution of diffusion coefficient. The ensemble average of the diffusion coefficient was calculated as $1.3~\mu m^2/s$, whereas the diffusion coefficient of the QDs in the bulk solution was estimated as $49~\mu m^2/s$ by the Einstein-Stokes equation. This indicated that the diffusion motion was suppressed by the walls of the column. Fig. 4b presents a histogram of the MSS slope. Approximately 70% of the trajectories were $0 < S_{\mathrm{MSS}} < 0.5$. This result supported the existence of suppressed diffusion.

The DPDs for the measured trajectories are shown in Fig. 5. At $\Delta r > 0.17~\mu m$, the DPD deviated from the Gaussian distribution. This deviation can be explained by the absorption and/or desorption of the QDs, as shown in Fig. 3. Although such deviations are observed in the trapped and hopping dynamics of particle motion in liposome diffusion in a nematic solution [43] and nanoparticle diffusion in a polymer solution, [44, 45] the deviation has not been reported for particle motion in a porous material to our knowledge.

For a more detailed analysis, the relative angles defined in Eq. 8 were calculated, as shown in Fig. 6. The relative angle distribution peaked at $\theta = 180°$ and was different from a uniform distribution corresponding to free diffusion motion. A QD diffusing inside a pore moves backward when it encounters a wall; thus, the distribution peaks at 180°. This phenomena is also observed in the particle motion near the dead ends in a cell.[41] This is another evidence that the diffusion motion inside the pore is suppressed.

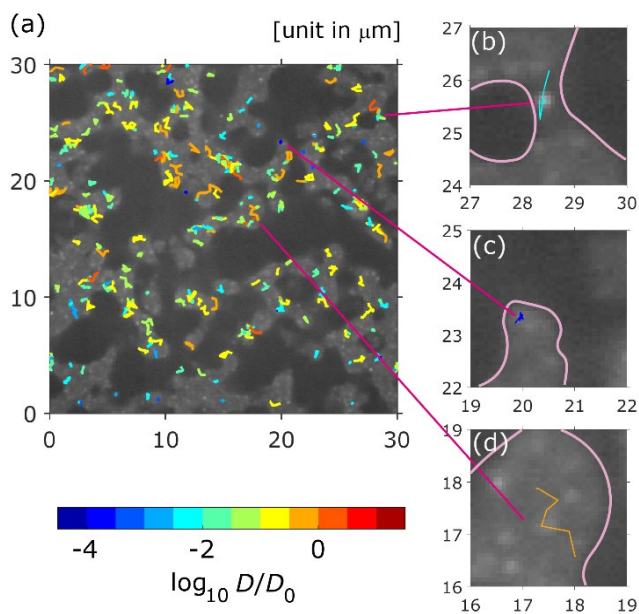

Fig. 3 Typical image of measured trajectories. Each trajectory is coloured by its diffusion coefficient $D$, where the diffusion coefficient is normalised by $D_0 = 1.0~\mu m^2/s$. (a) Trajectory distribution in an area of 30 μm × 30 μm. (b) QD trajectory with absorption and desorption. (c) Trajectory of an absorbed QD. (d) trajectory of freely diffusing QD. Pore walls are highlighted with pink lines in (b), (c), and (d).

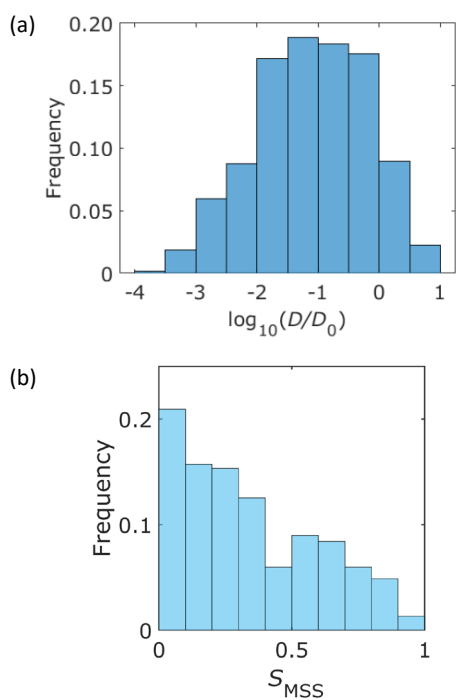

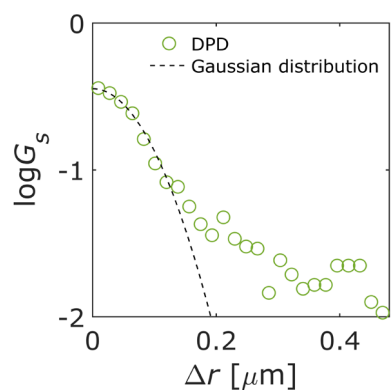

Fig. 5  Displacement probability distribution (DPD). Dashed line represents the Gaussian distribution.

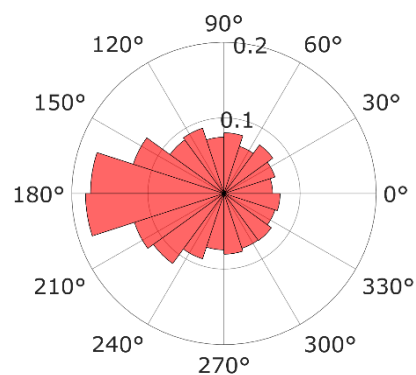

Fig. 4  Statistical analysis of measured trajectories. (a) Histogram of diffusion coefficient, which is normalized by $D_0 = 1.0\ \mathrm{\mu m^2/s}$. (b) Histogram of MSS slope $S_{\mathrm{MSS}}$.

Fig. 6  Relative angle distribution.

## Conclusions

In this study, we investigated nano-particle motions inside a porous monolithic silica column using the single-particle tracking (SPT) method. The ensemble-averaged diffusion coefficient of nano-particle was suppressed to approximately 1/40 of that of the particle in the bulk solution. The moment scaling spectra analysis indicated that approximately 70% of the particles were under suppressed diffusion. We calculated the displacement probability distribution (DPD) and found that the deviation from the Gaussian distribution at the tail. This deviation was induced by the absorption and/or desorption of the particles at the wall. These motions were also observed in the SPT trajectories. Moreover, the probability of the particle moving backward was approximately twice that of the particle moving forward. These results indicate that particle motion inside porous materials is highly complex; thus, the SPT method is a powerful tool for investigating the physicochemical properties of porous materials.

## Author Contributions

Yusaku Abe: Conceptualization, Data curation, Formal analysis, Investigation, Methodology, Software, Validation, Visualization, Writing – original draft, Writing – review & editing.

Naoki Tomioka: Investigation, Methodology

Yu Matsuda: Conceptualization, Data curation, Funding acquisition, Investigation, Methodology, Project administration, Supervision, Writing – original draft, Writing – review & editing

## Conflicts of interest

There are no conflicts to declare.

## Acknowledgements

The authors express their gratitude to R. Numajiri and H. Imai for their assistance with the SPT measurements. This work was partly supported by the JSPS, Japan Grant-in-Aid for Scientific Research (B), Nos. 19H02086 and 22H01421.

## References

1. K. Ishizaki, S. Komarneni and M. Nanko, *Porous Materials: Process technology and applications*, Springer US, 2013.
2. W. Ehlers and J. Bluhm, *Porous Media: Theory, Experiments and Numerical Applications*, Springer, 2002.
3. A. Uthaman, S. Thomas, T. Li and H. Maria, *Advanced Functional Porous Materials: From Macro to Nano Scale Lengths*, Springer International Publishing, 2021.
4. J. Shi, H. Du, Z. Chen and S. Lei, *Appl. Therm. Eng.*, 2023, **219**.

5. S. J. Mooney, I. M. Young, R. J. Heck and S. Peth, *X-ray Imaging of the Soil Porous Architecture*, Springer International Publishing, 2022.
6. E. F. Médici and A. D. Otero, *Album of Porous Media: Structure and Dynamics*, Springer International Publishing, 2023.
7. C. Schlumberger and M. Thommes, *Advanced Materials Interfaces*, 2021, **8**.
8. D. Schneider, D. Mehlhorn, P. Zeigermann, J. Karger and R. Valiullin, *Chem. Soc. Rev.*, 2016, **45**, 3439-3467.
9. P. S. Burada, P. Hanggi, F. Marchesoni, G. Schmid and P. Talkner, *Chemphyschem*, 2009, **10**, 45-54.
10. D. S. Grebenkov and G. Oshanin, *Phys. Chem. Chem. Phys.*, 2017, **19**, 2723-2739.
11. X. Yang, C. Liu, Y. Li, F. Marchesoni, P. Hanggi and H. P. Zhang, *Proc Natl Acad Sci U S A*, 2017, **114**, 9564-9569.
12. S. K. Ghosh, A. G. Cherstvy and R. Metzler, *Phys. Chem. Chem. Phys.*, 2015, **17**, 1847-1858.
13. V. Beschieru, B. Rathke and S. Will, *Microporous Mesoporous Mater.*, 2009, **125**, 63-69.
14. I. Teraoka, K. H. Langley and F. E. Karasz, *Macromolecules*, 1993, **26**, 287-297.
15. M. Kaasalainen, V. Aseyev, E. von Haartman, D. Ş. Karaman, E. Mäkilä, H. Tenhu, J. Rosenholm and J. Salonen, *Nanoscale Research Letters*, 2017, **12**, 74.
16. S. M. Mahurin, S. Dai and M. D. Barnes, *The Journal of Physical Chemistry B*, 2003, **107**, 13336-13340.
17. A. Masuda, K. Ushida and T. Okamoto, *Physical Review E*, 2005, **72**.
18. B. Dong, N. Mansour, T.-X. Huang, W. Huang and N. Fang, *Chem. Soc. Rev.*, 2021, **50**, 6483-6506.
19. J. J. E. Maris, D. Fu, F. Meirer and B. M. Weckhuysen, *Adsorption*, 2021, **27**, 423-452.
20. H. Qian, M. P. Sheetz and E. L. Elson, *Biophys. J.*, 1991, **60**, 910-921.
21. A. Kusumi, Y. Sako and M. Yamamoto, *Biophys. J.*, 1993, **65**, 2021-2040.
22. M. J. Saxton, *Biophys. J.*, 1997, **72**.
23. A. Zürner, J. Kirstein, M. Doblinger, C. Brauchle and T. Bein, *Nature*, 2007, **450**, 705-708.
24. J. Kirstein, B. Platschek, C. Jung, R. Brown, T. Bein and C. Brauchle, *Nat Mater*, 2007, **6**, 303-310.
25. B. Rühle, M. Davies, T. Lebold, C. B. uchle and T. Bein, *ACS Nano*, 2012, **6**, 1948-1960.
26. Y. Liao, S. K. Yang, K. Koh, A. J. Matzger and J. S. Biteen, *Nano Lett.*, 2012, **12**, 3080-3085.
27. R. Kumarasinghe, T. Ito and D. A. Higgins, *Anal. Chem.*, 2020, **92**, 1416-1423.
28. J. T. Cooper, E. M. Peterson and J. M. Harris, *Anal. Chem.*, 2013, DOI: 10.1021/ac402251r.
29. F. C. Hendriks, F. Meirer, A. V. Kubarev, Z. Ristanovic, M. B. J. Roeffaers, E. T. C. Vogt, P. C. A. Bruijnincx and B. M. Weckhuysen, *J. Am. Chem. Soc.*, 2017, **139**, 13632-13635.
30. K. Nakanishi, R. Takahashi, T. Nagakane, K. Kitayama, N. Koheiya, H. Shikata and N. Soga, *J. Sol-Gel Sci. Technol.*, 2000, **17**, 191-210.
31. R. Iwao, H. Yamaguchi, T. Niimi and Y. Matsuda, *Physica A: Statistical Mechanics and its Applications*, 2021, **565**.
32. I. F. Sbalzarini and P. Koumoutsakos, *J Struct Biol*, 2005, **151**, 182-195.
33. I. F. Sbalzarini and P. Koumoutsakos, MOSAIC Software Repository, http://mosaic.mpi-cbg.de/?q=downloads#).
34. N. Chenouard, I. Smal, F. de Chaumont, M. Maska, I. F. Sbalzarini, Y. Gong, J. Cardinale, C. Carthel, S. Coraluppi, M. Winter, A. R. Cohen, W. J. Godinez, K. Rohr, Y. Kalaidzidis, L. Liang, J. Duncan, H. Shen, Y. Xu, K. E. Magnusson, J. Jalden, H. M. Blau, P. Paul-Gilloteaux, P. Roudot, C. Kervrann, F. Waharte, J. Y. Tinevez, S. L. Shorte, J. Willemse, K. Celler, G. P. van Wezel, H. W. Dan, Y. S. Tsai, C. Ortiz de Solorzano, J. C. Olivo-Marin and E. Meijering, *Nat Methods*, 2014, **11**, 281-289.
35. J. Schindelin, I. Arganda-Carreras, E. Frise, V. Kaynig, M. Longair, T. Pietzsch, S. Preibisch, C. Rueden, S. Saalfeld, B. Schmid, J. Y. Tinevez, D. J. White, V. Hartenstein, K. Eliceiri, P. Tomancak and A. Cardona, *Nat Methods*, 2012, **9**, 676-682.
36. R. Ferrari, A. J. Manfroi and W. R. Young, *Physica D: Nonlinear Phenomena*, 2001, **154**, 111-137.
37. A. M. Michalak and P. K. Kitanidis, *Stochastic Environmental Research and Risk Assessment*, 2005, **19**, 8-23.
38. M. V. Chubynsky and G. W. Slater, *Phys. Rev. Lett.*, 2014, **113**, 098302.
39. R. Metzler, J. H. Jeon, A. G. Cherstvy and E. Barkai, *Phys. Chem. Chem. Phys.*, 2014, **16**, 24128-24164.
40. S. Burov, S. M. Tabei, T. Huynh, M. P. Murrell, L. H. Philipson, S. A. Rice, M. L. Gardel, N. F. Scherer and A. R. Dinner, *Proc Natl Acad Sci U S A*, 2013, **110**, 19689-19694.
41. I. Izeddin, V. Recamier, L. Bosanac, Cisse, II, L. Boudarene, C. Dugast-Darzacq, F. Proux, O. Benichou, R. Voituriez, O. Bensaude, M. Dahan and X. Darzacq, *Elife*, 2014, **3**.
42. Y. Matsuda, I. Hanasaki, R. Iwao, H. Yamaguchi and T. Niimi, *Anal. Chem.*, 2016, **88**, 4502-4507.
43. B. Wang, J. Kuo, S. C. Bae and S. Granick, *Nature Materials*, 2012, **11**, 481-485.
44. C. Xue, X. Zheng, K. Chen, Y. Tian and G. Hu, *J Phys Chem Lett*, 2016, **7**, 514-519.
45. B. Wang, S. M. Anthony, S. C. Bae and S. Granick, *Proceedings of the National Academy of Sciences*, 2009, **106**, 15160-15164.